\documentclass{aa}
\def\folio{\ifnum\pageno=1\nopagenumbers\else\number\pageno\fi}

\def\lax    {\ifmmode{_<\atop^{\sim}}\else{${_<\atop^{\sim}}$}\fi}
\def\gax    {\ifmmode{_>\atop^{\sim}}\else{${_>\atop^{\sim}}$}\fi}
\newbox\grsign      \setbox\grsign=\hbox{$>$} 
\newdimen\grdimen   \grdimen=\ht\grsign
\newbox\simgreatbox \setbox\simgreatbox=\hbox{\raise.5ex\hbox{$>$}\llap
                        {\lower.5ex\hbox{$\sim$}}}\ht1=\grdimen\dp1=0pt
\newbox\simlessbox  \setbox\simlessbox =\hbox{\raise.5ex\hbox{$<$}\llap
                        {\lower.5ex\hbox{$\sim$}}}\ht2=\grdimen\dp2=0pt

\newbox\grsign \setbox\grsign=\hbox{$>$} \newdimen\grdimen \grdimen=\ht\grsign
\newbox\laxbox \newbox\gaxbox
\setbox\gaxbox=\hbox{\raise.5ex\hbox{$>$}\llap
     {\lower.5ex\hbox{$\sim$}}}\ht1=\grdimen\dp1=0pt
\setbox\laxbox=\hbox{\raise.5ex\hbox{$<$}\llap
     {\lower.5ex\hbox{$\sim$}}}\ht2=\grdimen\dp2=0pt
\def\gax{\mathrel{\copy\gaxbox}}
\def\lax{\mathrel{\copy\laxbox}}
\def\boxit#1    {\vbox{\hrule\hbox{\vrule\kern3pt
                  \vbox{\kern3pt#1\kern3pt}\kern3pt\vrule}\hrule}}
\def\h      {\ifmmode{^{\rm h}}\else{$^{\rm h}$}\fi}
\def\m      {\ifmmode{^{\rm m}}\else{$^{\rm m}$}\fi}
\def\s      {\ifmmode{^{\rm s}}\else{$^{\rm s}$}\fi}
\def\decas    {\ifmmode{{\rlap.}{''}}\else{${\rlap.}{''}$}\fi}
\def\mum     {\ifmmode{\mu{\rm m}}\else{$\mu{\rm m}$}\fi}
\def\s      {\ifmmode{^{\rm s}}\else{$^{\rm s}$}\fi}
\def\deg      {\ifmmode{^{\circ}}\else{$^{\circ}$}\fi}
\def\as     {\ifmmode {\rlap.}$\,$''$\,$\! \else ${\rlap.}$\,$''$\,$\!$\fi}
\def\decsec  {\ifmmode {\rlap.}$\,$^{s}$\,$\! \else ${\rlap.}$\,$^{s}$\,$\!$\fi}\def\decs  {\ifmmode {\rlap.}$\,$^{s}$\,$\! \else ${\rlap.}$\,$^{s}$\,$\!$\fi}

\def\kms    {\ifmmode{{\rm km~s}^{-1}}\else{km~s$^{-1}$}\fi}

\def\Mspy   {\ifmmode {M_{\odot} {\rm yr}^{-1}} \else $M_{\odot}$~yr$^{-1}$\fi}
\def\Mdot   {\ifmmode {\dot M} \else $\dot M$\fi}
\def\mhd    {\ifmmode {n_{{\rm H}_2}} \else $n_{{\rm H}_2}$\fi}
\def\mhcd   {\ifmmode {N_{{\rm H}_2}} \else $N_{{\rm H}_2}$\fi}

\def\El      {\ifmmode{E_{\ell}}\else{$E_{\ell}$}\fi}
\def\beam    {\ifmmode{\theta_{\rm B}}\else{$\theta_{\rm B}$}\fi}
\def\mjyb   {\ifmmode {{\rm mJy~beam}^{-1}} \else{mJy~beam$^{-1}$}\fi}
\def\mujyb   {\ifmmode {\mu{\rm Jy~beam}^{-1}} \else{$\mu$Jy~beam$^{-1}$}\fi}
\def\Trot   {\ifmmode{T_{\rm rot}}\else$T_{\rm rot}$\fi}    
    
\def\Teff   {\ifmmode{T_{\rm eff}}\else$T_{\rm eff}$\fi}

\def\ITRS   {\ifmmode{\smallint {\rm T}_{R}^{*}dv}\else{$\smallint 
{\rm T}_{R}^{*}dv$}\fi}
\def\ITRS   {\ifmmode{\smallint {\rm T}_{R}^{*}dv}\else{$\smallint 
{\rm T}_{R}^{*}dv$}\fi}
\def\ITAS   {\ifmmode{\smallint {\rm T}_{A}^{*}dv}\else{$\smallint 
{\rm T}_{A}^{*}dv$}\fi}

\def\HII        {H~{\eightpt II}}

\def\lefttitle#1  {\noindent \hangindent=18.0pt \hangafter=1 {#1} \par}
\def\vol#1  {{\bf {#1}{\rm,}\ }}
\font\tenssb=cmssbx10
\font\tenbf=cmbx10
\font\sevenbf=cmbx8
\font\fivebf=cmbx6
\def\unetdemi    {\smallskipamount=6pt plus2pt minus2pt
                  \medskipamount=12pt plus4pt minus4pt
                  \bigskipamount=24pt plus8pt minus8pt
                  \normalbaselineskip=16pt plus0pt minus0pt
                  \normallineskip=2pt
                  \normallineskiplimit=0pt
                  \jot=6pt
                  {\def\smallskip {\vskip\smallskipamount}}
                  {\def\medskip   {\vskip\medskipamount}}
                  {\def\bigskip   {\vskip\bigskipamount}}
                  {\setbox\strutbox=\hbox{\vrule 
                    height17.0pt depth7.0pt width 0pt}}
                  \parskip 12.0pt
                  \normalbaselines}
\def\smallerspace {\smallskipamount=3pt plus0pt minus0pt
                  \medskipamount=6pt plus0pt minus0pt
                  \bigskipamount=10.5pt plus0pt minus0pt
                  \normalbaselineskip=10.5pt plus0pt minus0pt
                  \normallineskip=1pt
                  \normallineskiplimit=0pt
                  \jot=3pt
                  {\def\smallskip {\vskip\smallskipamount}}
                  {\def\medskip   {\vskip\medskipamount}}
                  {\def\bigskip   {\vskip\bigskipamount}}
                  {\setbox\strutbox=\hbox{\vrule 
                    height8.5pt depth3.5pt width 0pt}}
                  \parskip 0pt
                  \normalbaselines}
\def\memospace    {\smallskipamount=4pt plus1pt minus1pt
                  \medskipamount=6pt plus2pt minus2pt
                  \bigskipamount=14pt plus6pt minus6pt
                  \normalbaselineskip=14pt plus0pt minus0pt
                  \normallineskip=1pt
                  \normallineskiplimit=0pt
                  \jot=4pt
                  {\def\smallskip {\vskip\smallskipamount}}
                  {\def\medskip   {\vskip\medskipamount}}
                  {\def\bigskip   {\vskip\bigskipamount}}
                  {\setbox\strutbox=\hbox{\vrule 
                    height17.0pt depth7.0pt width 0pt}}
                  \parskip 2.0pt
                  \normalbaselines}
\def\memowidespace    {\smallskipamount=5pt plus1pt minus1pt
                  \medskipamount=7.5pt plus2pt minus2pt
                  \bigskipamount=17.5pt plus6pt minus6pt
                  \normalbaselineskip=17.0pt plus0pt minus0pt
                  \normallineskip=1.25pt
                  \normallineskiplimit=0pt
                  \jot=5pt
                  {\def\smallskip {\vskip\smallskipamount}}
                  {\def\medskip   {\vskip\medskipamount}}
                  {\def\bigskip   {\vskip\bigskipamount}}
                  {\setbox\strutbox=\hbox{\vrule 
                    height21.25pt depth8.75pt width 0pt}}
                  \parskip 2.5pt
                  \normalbaselines}
\usepackage{graphics,epsf}

\usepackage{natbib}
\usepackage{ulem}
      \def\new#1 {{\bf #1 }}
      \def\cut#1 {\sout{#1} }

\def\TWCO {$\mathrm{^{12}CO}$} 
\def\CEIO {$\mathrm{C^{18}O}$} 

\def\percc {$\mathrm{cm^{-3}}$} 
\def\cmsq  {$\hbox{{\rm cm}}^{-2}$}    
\def\FORM {$\mathrm{H_2CO}$} 
\def\METH {$\mathrm{CH_3OH}$} 
\def\NTHP {$\mathrm{N_2H^+}$} 
\def\Lsol {$\hbox{L}_\odot$}
\def\Msol {$\hbox{M}_\odot$}
\def\HII  {H{\sc ii}}

\begin{document}

\title{Revealing the environs of the remarkable southern hot core G327.3--0.6}
\author{F. Wyrowski\inst{1}, K. M. Menten\inst{1}, P. Schilke\inst{1}, 
        S. Thorwirth\inst{1}, R. G\"usten\inst{1},
\and
P. Bergman\inst{2,3}
}

\offprints{F. Wyrowski}

\institute{Max-Planck-Institut f\"ur Radioastronomie, Auf dem H\"ugel
69, D-53121 Bonn, Germany
\and 
European Southern Observatory, Alonso de Cordova 3107, Vitacura Casilla 19001, Santiago
19, Chile
\and
Onsala Space Observatory, Chalmers University of
Technology, SE-439 92 Onsala, Sweden}
\date{Received / Accepted}
\titlerunning{The environs of the hot core G327.3--0.6}
\authorrunning{Wyrowski et al.}


\abstract
{} 
{We present a submm study of the massive hot core G327.3--0.6 that
  constrains its physical parameters and environment.}
{The APEX telescope was used to image CO and \NTHP\ emission, to
  observe lines from other molecules toward a hot and a cold molecular
  core, and to measure the continuum flux density of the hot core.}
{In the \CEIO\ $J=3 - 2$ line, two clumps were found, one associated
  with the \HII\ region G327.3--0.5 and the other associated with the
  hot core. An additional cold clump is found 30\arcsec\ (0.4 pc)
  northeast of the hot core in bright \NTHP\ emission. From the the
  continuum data, we calculate a mass of 420~\Msol\ and a size of
  0.1~pc for the hot core. A new, more accurate position of the hot
  core is reported, which allows the association of the core with a
  bright mid-infrared source. The luminosity of the hot core is
  estimated to be between $5$ and $15\times 10^4$~\Lsol.}
{This study revealed several different evolutionary stages of massive
  star formation in the G327.3--0.6 region.}


\maketitle

\section{\label{intro}Introduction}

There is no generally accepted evolutionary scheme for high-mass star
formation yet, in contrast to the detailed framework of CLASSes that
exists for the early evolution of low-mass stars. Observationally,
hot molecular cores represent an early evolutionary stage during the
formation of massive stars.
One of the most prominent hot molecular cores in the southern
celestial hemisphere is the core associated with the \HII\ region
G327.3--0.6 at a kinematical distance of 2.9 kpc (Bergman 1992). It is
associated with prominent H$_2$O, OH, and CH$_3$OH masers, and its chemistry
has been studied in two papers, %
one reporting  ethylene oxide and acetaldehyde
observations (Nummelin et al.\ 1998), while the other  
investigates
the chemical inventory of this source (Gibb et al.\ 2000). The source is remarkable for
its exceptionally  rich molecular line spectra with relatively narrow, 
well-behaved (Gaussian) line profiles. This  reduces line blending and
makes them easier to interpret than spectra of many other very 
line-rich sources, such as 
Sgr B2.
It is therefore surprising that almost
nothing is known about its environs. Only Bergman (1992) reports some SEST
maps that reveal  two adjacent dense cores in this molecular
cloud: one relatively cold ($T_{\rm kin} \sim 30$~K) cloud core and one hot
(T = 100--200~K) core. Hence, this region offers the possibility to study
cores that have formed from the same parental cloud, but that are in
different stages of evolution.

Since the G327.3--0.6 hot core has the potential of becoming a
southern hemisphere hot core template for upcoming observatories like
ALMA, we revisited this source and its environment with the recently
commissioned APEX telescope (see G\"usten et al., this volume).

\section{\label{obs}Observations}

The observations were done with the Atacama
Pathfinder Experiment (APEX\footnote{This publication is based on data
  acquired with the Atacama Pathfinder Experiment (APEX). APEX is a
  collaboration between the Max-Planck-Institut f\"ur Radioastronomie,
  the European Southern Observatory, and the Onsala Space
  Observatory.}).  The frontends
used were the facility 345~GHz and the MPIfR dual channel (460 and 810
GHz) FLASH receiver (Risacher et al.\ 2006, Heyminck et al.\ 2006, this
volume). As backends, the MPIfR pocket backend (PBE) for the continuum
and the MPIfR Fast Fourier Transform Spectrometer (FFTS, Klein et al.\ 
2006, this volume) for the line
observations were used. With the 345~GHz receiver, several arcmin$^2$-sized 
maps of \TWCO\ and \CEIO\ (3--2) were observed using the On-The-Fly (OTF)
observing technique. \NTHP\ (3--2) was observed with raster mapping
and the (4--3) and (5--4) transitions with single pointings towards
the peak of the (3--2) emission. All observations were done in
position switching mode. The continuum observations with the 345~GHz
and FLASH receivers were done as cross scans over the source 
using the continuum PBE.  We used the hot core position
($\alpha_{J2000},\delta_{J2000}$)=(15:53:08.78,--54:37:01.20) from
Bergman (1992) as the center of the maps.

\section{\label{results}Results}
\subsection{CO mapping}

\begin{figure}
\begin{center}
  \epsfxsize=7cm \rotatebox{-90}{\epsfbox{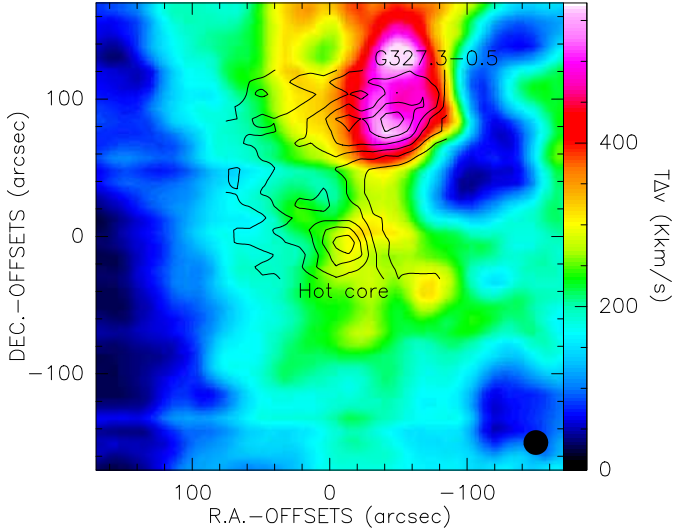}}
     \caption{ APEX \TWCO\ (color) and \CEIO\ (3--2) (contours) 
               images. The contour levels start at 23.6 with steps of 
               15.6 K\kms.
       \label{fig:apex_co} }
\end{center}
\end{figure} 

The large CO (3-2) map in Fig.~\ref{fig:apex_co} shows
widespread $J=3-2$ emission. 
The strongest emission, indicating elevated temperatures in this
optically thick line, is found in the north, and is associated with the
strong \HII\ region G327.3--0.5 (Goss \& Shaver 1970) and a complex of
infrared sources around IRS3 (Epchtein \& Lepine 1981).  Lines 
in this
region have temperatures going up to 70~K, with widths of 10~km~s$^{-1}$
and typical velocities of $-48$~\kms. This CO
emission likely traces the hot surface of a Photon Dominated Region (PDR)
around G327.3--0.5.

There is an extension in the CO emission to the south where the hot
core is located, but it is weaker by a factor of 2. Typical line
temperatures and velocities in this region are 30~K and --46~\kms,
respectively.  Close, but offset from the hot core, the line profiles
show weak evidence for outflow motions.

To the east, the CO emission vanishes rapidly, and to the west of 
the strong northern peak, it shows a conspicuous hourglass shaped
``hole'' where the CO emission is strongly reduced.  This region
might have been excavated by strong winds of an older generation of
luminous stars, although with a SIMBAD search within 1\arcmin, 
no objects were found towards the center of
the two evacuated lobes. In
the GLIMPSE images, shell-like structures are found around the holes,
as is expected for swept up material around stellar wind/radation
cavities (Churchwell et al.\ 2004). Another less striking hole in CO and MIR emission is at
(5\arcsec, 140\arcsec).

Toward the northern peak and the hot core in the south, the line profiles
show self absorption hinting toward a more complicated structure/layering
with excitation gradients in the column density peaks seen
in \CEIO\ (contours in Fig.~\ref{fig:apex_co}).  The hot core is clearly
identified as a \CEIO\ peak, as is 
a ridge of strong \CEIO\ emission associated
with the \HII\ region G327.3--0.5. The peak within the ridge is at the
position of one of the two strong \TWCO\ (3--2) emission maxima.
Using the \CEIO\ (1--0) and (2--1) line intensities, as
measured by Bergman (1992), together with the new (3--2) measurement, we
determine a column density of $6\times 10^{16}$\cmsq\ and an
excitation temperature of 50 K, which is somewhat higher than the
temperature measured in \TWCO, and is likely to be caused by a combination of increasing
temperature to the inner part of the hot core and clumping within the
beam. Using a \CEIO\ abundance of $2\times10^{-7}$ (Frerking et al.\
1982), the mass in the beam is then 510~\Msol. With a size of
25\arcsec\ (after subtracting the large scale \CEIO\ emission) for the
clump harboring the hot core, a virial mass can be estimated to be
1330~\Msol, using the procedure described by Cesaroni et al.\ (1994).

\subsection{CH$_3$OH results}

A strong methanol line was observed simultaneously with \CEIO\ 
(Table~\ref{tab:lineparam} and Fig.~\ref{fig:apex_n2h+}).  The hot core is very prominent in
methanol, but within the region mapped, it is the only methanol peak.
The northern \CEIO\ peak was not detected in this molecule, which is
surprising given its high temperatures and column densities. 
This might either be due to lower densities in the northern region so that the
methanol transitions cannot be sufficiently excited or due to low methanol abundances
in PDRs (Jansen et al.\ 1995).   
The peak of the
emission is slightly offset by 5\arcsec\ from the one found by Bergman
et al.\ (1992). The deconvolved size of the emission is 26x7\arcsec,
with P.A.=60\deg\ to the northeast.

\subsection{Continuum}


Continuum cross-scans of the hot core found the peak emission at
(--6\arcsec, --6\arcsec), where the \METH\ also peaks. This is also
consistent with the maps of complex molecules observed by Gibb et al.\
(2000).  
Fit results from averaged cross-scans at different frequencies are 
 given in Table~\ref{tab:cont}.
The deconvolved size of the dust emission of the hot core is
 7\arcsec\ or 0.1~pc, and is consistent at all three frequencies.  The submm
flux density measured towards the hot core is almost entirely due to
thermal dust emission and allows a mass estimate of the hot core.
Using the flux density at 850~$\mum$, a dust emissivity of 1.8 from
Ossenkopf \& Henning (1994, model 5), and a temperature of 100~K
results in 420~\Msol\ for the core. For a Hildebrand (1983) opacity 
with $\beta=2$, the mass is 950~\Msol.

\begin{table}
\caption{APEX continuum results}
\begin{tabular}{ccccc}
\hline
\hline
Frequency & $T_{\rm MB}$ & Widths    & Dec. Widths & Flux density \\
  (GHz)   &      (K)     & (\arcsec) &  (\arcsec)  &  (Jy)        \\
\hline
351 & 1.8 & 19.1 & 7.7 &  66 \\
464 & 3.0 & 14.8 & 6.6 & 117 \\
813 & 8.3 & 10.8 & 7.7 & 525 \\
\hline
\end{tabular}
\label{tab:cont}
\end{table}

\subsection{Spectral energy distribution}

\begin{figure}
\begin{center}
      \epsfxsize=6cm
      \rotatebox{-90}{\epsfbox{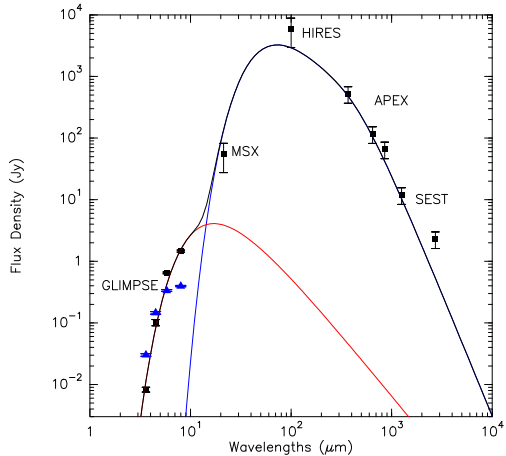}}
     \caption{Spectral energy distribution of the G327.3--0.6 hot core fit
              with a 2-component greybody fit. GLIMPSE fluxes of the
              source embedded in the cold core are marked as triangles.
              \label{fig:sed} }
\end{center}
\end{figure}

With the improved position from our \METH\ and continuum observations,
a clear association of the hot core with GLIMPSE point sources becomes
possible (see Fig.~\ref{fig:apex_n2h+}). At the position of the hot
core, a bright 8~$\mum$ source is found that obeys the color and
brightness criteria suggested by Ellingsen (2006) for massive star-forming 
cores harboring methanol masers. Only a few arcseconds off,
another weaker source with high extinction is found, likely embedded
in the hot core as well.  In Fig.~\ref{fig:sed}, the spectral energy
distribution is shown towards the hot core. Additional millimeter flux
densities are taken from Bergman (1992).  Brand et al.\ (1984)
estimate a 1.4 GHz flux of 1~Jy of what they call the southern
extension of the \HII\ region G326.3--0.5. This free-free flux and more
extended dust emission might contribute to the excess emission seen in
the 45\arcsec\ SEST beam at 3~mm.  Far infrared data points are the most
important means of constraining the total luminosity of
embedded young stellar objects. G327.3--0.6 itself is not associated with an IRAS point
source, likely due to confusion or blending with the strong source
IRAS 15492-5426. We checked 100~$\mum$ HIRES images of the source and
found that it breaks up into two sources; one is close in position to the
hot core, and the flux of this source is given in Fig.~\ref{fig:sed}.
Also, the hot core is not associated with any MSX source. The closest
MSX source is the extended feature 10\arcsec\ to the west of the hot core,
we therefore estimated the 20~$\mum$ MSX flux of the hot core directly
from MSX images. We take the high uncertainty in the 20 and 100~$\mum$
fluxes, due to the confusion with other sources with 50\% error bars,
into account. In Fig.~\ref{fig:sed}, a simple 2-component fit of all
the flux densities is shown, with a greybody at 70~K fitting the 
mm/submm/FIR part and a blackbody at 300~K fitting the MIR emission
of the GLIMPSE source. Given the uncertainties of the FIR fluxes,
we estimate a luminosity between $5-15\times10^4$~\Lsol\ for the
hot molecular core.

\begin{table*}
\begin{center}
\caption{Line parameters towards molecular cores in the G327 region.
         For \NTHP\ $T_{\rm peak}$ was determined from Gaussian fits
          and the line widths from simultaneously fitting of all HFS components
          assuming the same excitation for all them, 
         resulting in optical depths between 1--2. We also accounted for the
         blend of the \FORM\ $4_{32}-3_{31}$ and $4_{31}-3_{30}$ lines.}
\begin{tabular}{lcccccccc}
\hline
\hline
Source & Transition & Frequency & $E_{\rm lower}$ & Offsets & Size & $T_{\rm peak}$  & $\Delta v$ & $v_{\rm LSR}$ \\
       &            &    (MHz)  & (K)             & (\arcsec,\arcsec) & (\arcsec) & (K) & (\kms)    & (\kms)    \\
\hline
North core& \CEIO\ 3--2 & 329330.600 & 15.8 & (-44,82)&      & 21.3 (0.7) & 4.9 (0.1) & -47.5 (0.1) \\
Hot core  & \CEIO\ 3--2 &            & & (-6,-6) & 25   & 15.8 (1.0) & 6.2 (0.2) & -44.4 (0.1) \\
 & \METH\ $7_{43}-6_{43}$  & 341415.625 & 63.7 &    & 26x7  &  9.3 (0.8) & 6.6 (0.6) & -45.3 (0.2)  \\
          & \NTHP\ 3--2 & 279511.701 & 13.4 &        &       &  7.9 (0.2) & 6.0 (1.5) & -45.3 (0.3) \\
Cold core & \NTHP\ 3--2 &            &      & (12,6) & 77x39 & 12.2 (0.2) & 4.8 (0.2) & -45.9 (0.1) \\
          & \NTHP\ 4--3 & 372672.509 & 26.8 & (12,6) &       & 11.4 (0.2) & 4.3 (0.1) & -45.8 (0.1) \\
          & \NTHP\ 5--4 & 465824.947 & 44.7 & (12,6) &       &  9.0 (0.4) & 4.4 (0.2) & -45.4 (0.1) \\
 & \FORM\ $4_{23}-3_{22}$ & 291237.781 & 68.1 & (12,6) &    & 2.3 (0.2) & 4.5 (0.2) & -45.5 (0.1) \\
 & \FORM\ $4_{32}-3_{31}$ & 291380.500 & 127.0& (12,6) &    & 1.3 (0.2) & 7.7 (0.4) & -45.2 (0.2) \\
\hline
\end{tabular}
\label{tab:lineparam}
\end{center}
\end{table*}



\subsection{N$_2$H$^+$ mapping}

\begin{figure}
\begin{center}
      \epsfxsize=7.5cm
      \epsfxsize=6cm
      \rotatebox{-90}{\epsfbox{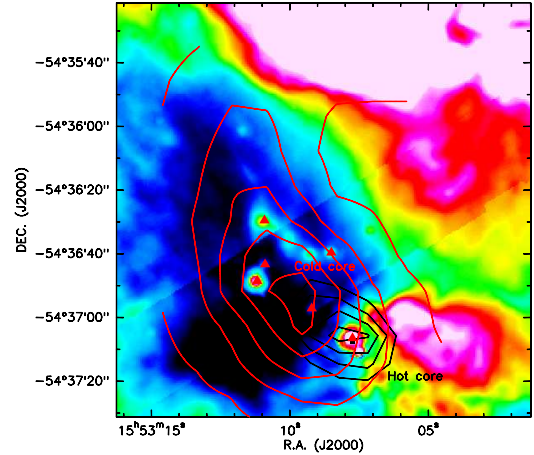}}
     \caption{APEX images in the \NTHP\ (3--2) and the \METH\ $7_{43}-6_{43}$ 
              lines (red and black contours)
              overlaid on the 8~$\mu$m GLIMPSE emission. Embedded
              GLIMPSE point sources with massive YSO characteristics
              are marked with triangles. \label{fig:apex_n2h+} }
\end{center}
\end{figure}  

\begin{figure}
\begin{center}
      \epsfxsize=6.cm
      \rotatebox{0}{\epsfbox{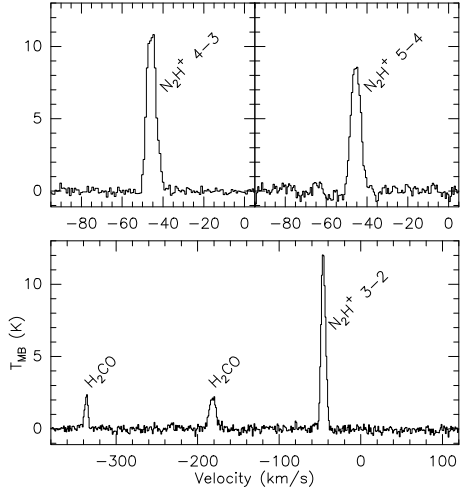}}
     \caption{ APEX \NTHP\ spectra observed toward the cold core at
             (12\arcsec, 6\arcsec).
             \label{fig:apex_n2h+_spectra} }

\end{center}
\end{figure}

Figure~\ref{fig:apex_n2h+} shows the 8~$\mum$ GLIMPSE image of the
region with contours of \METH\ and \NTHP\ (3--2)
overlaid. The hot core seen in methanol is situated at the edge of an
arcmin scale mid-infrared dark cloud, which is traced 
well in
\NTHP. The \NTHP\ peak seems to trace a colder clump in an earlier
evolutionary state. Interestingly, even the cold clump already harbors
a hotter component, as is evident from the simultaneously
detected \FORM\ lines (Fig.~\ref{fig:apex_n2h+_spectra}) and the
highly reddened GLIMPSE sources (blue markers in Fig.~\ref{fig:sed}).
The total size of the \NTHP\ clump is 1 pc. 

The high and comparable peak temperatures of the 3 observed \NTHP\
lines (see Table~\ref{tab:lineparam}) suggest that they are optically
thick. We used RADEX
on-line\footnote{http://www.strw.leidenuniv.nl/$\sim$moldata/radex.php} to
investigate the required excitation conditions for the observed
intensities. Similar excitation temperatures for both lines can only
be reached for densities $>10^6$~\percc\ unless the \NTHP\
abundances are higher than $3\times10^{-9}$. With a kinetic temperature
of 18~K, the line intensities can be reproduced, but still require
$\rm H_2$ column densities of the order of $10^{24}$~\cmsq.
The high column density, together with weak \CEIO, suggests that \CEIO\
is depleted in the cold clump, and maybe even in the outer envelope of the
hot core clump, which would explain the discrepancy between \CEIO\ and
virial mass.
Using Cesaroni et al.\ (1994) again, the virial mass of the 60\arcsec\
\NTHP\ clump is 1000~\Msol.

\section{\label{conclusions}Conclusions}

A detailed view onto the environs of G327.3--0.6 is another example
of the complexity of massive star-forming regions, since
many observational phenomena occur simultaneously and can only
be separated with adequate angular resolution. In what we think
is the order of decreasing evolutionary stages, the phenomena in this region are: 
\begin{itemize}
\item suspicious holes
in the CO emission 
possibly excavated by B-star winds, 
\item the bright \HII\ region G327.3--0.5 to the north associated with a luminous PDR seen
in CO, 
\item an ultracompact \HII\ region 2 arcmin (1.7 pc) 
south, seen as
a southern extensions of the old cm continuum maps and as bright
extended MIR emission by GLIMPSE,
\item 5--10 arcsec (0.07 -- 0.14 pc) 
offset from the UC \HII\ region, the hot molecular core
harboring several MIR point sources, and
\item further east of the hot core, a dense and cold clump with an
embedded massive young stellar object.
\end{itemize}

The mass and luminosity of the hot core of about 500~\Msol\ and
$10^5$~\Lsol\ are comparable to ``classical'' luminous hot cores like
G10.47+0.03 and G31.41+0.31 (Hatchell et al.\ 2000), but for further studies
the advantage of G327.3--0.6
compared to those regions  will be that its distance
is a factor of 2 smaller.


\end{document}